\documentclass[aps,floatfix,12pt]{revtex4-2}

\usepackage{graphicx}
\usepackage[utf8]{inputenc}
\usepackage{amsmath,amsfonts}
\usepackage[colorlinks=true,citecolor=blue]{hyperref}

\begin{document}

\author{Lorenz Meyer}
\author{Nicolas Néel}
\author{Jörg Kröger}\email{joerg.kroeger@tu-ilmenau.de}
\affiliation{Institut für Physik, Technische Universität Ilmenau, D-98693 Ilmenau, Germany}

\title{Single-enantiomer spin polarisers in superconducting junctions}

\begin{abstract}
Chiral matter acting as a spin-selective device in biased electron transport is attracting attention for the quantum-technological design of miniaturized electronics.
To date, however, experimental reports on spin selectivity are not conclusive.  
The magnetoresistance in electron transport measurements observed for chiral materials on ferromagnets upon magnetisation reversal is proposed to result from electrostatic rather than from the sought-after chiral effects.
Recent break junction studies even question the spin-dependent electron flow across single chiral molecules.
Here, we avoid ferromagnetic electrodes and magnetisation reversal to provide unambiguous experimental evidence for the chirality-induced spin selectivity effect of single enantiomers.
Functionalising the superconducting tip of a scanning tunnelling microscope with a manganese atom cluster gives rise to Yu-Shiba-Rusinov resonances that serve as spin-sensitive probes of the tunnelling current in junctions of single heptahelicene molecules adsorbed on a crystalline lead surface.
Our key finding is the dependence of the signal strength of these states in spectroscopy of the differential conductance on the handedness of the molecule.
The experiments unveil the role of the enantiomers as spin polarisers and the irrelevance of electrostatics in the chosen model system.
\end{abstract}

\maketitle

The chirality-induced spin selectivity (CISS) effect was first reported from photoelectron spectroscopy of Langmuir-Blodgett films \cite{science_283_814}, where a preferential photoelectron spin state was observed for a given chiral domain of the film.
This pioneering work stimulated substantial experimental and theoretical studies, which have thoroughly been reviewed \cite{chemrev_124_1950,natrevchem_3_250,jpcm_37_113003}.
Despite its importance for fundamental physics and nanotechnological applications, the origin of the CISS effect, that is, the understanding of the chirality-induced locking of electron spin and momentum, is strongly debated and has remained elusive to date \cite{rmp_92_035001,acsnano_19_37484}.
This dilemma is partly due to ambiguities in electron transport experiments performed for chiral matter, which represent a major research direction devoted to the CISS effect.
In these studies, chiral molecules are attached to ferromagnetic electrodes, and changes in the magnetoresistance have been reported as the response to chirality change or to reversing the magnetisation of the ferromagnet \cite{acie_59_14671,jacs_143_7189,small_19_2302714,small_20_2308233}.
Recently, the interpretation of these observations in terms of the CISS effect has been questioned.
Rather, motivated by experimental observations \cite{jacs_141_3863,jpcl_11_1550,nl_23_8280}, a work function change upon magnetisation reversal has been proposed to feign the CISS effect \cite{acsnano_18_6028}.
Likewise, changes in the electric dipole of the chiral molecules can influence the CISS effect without altering the chirality \cite{natcommun_7_10744,chemsci_15_14905,jacs_147_36453}.
The alarming consequence of these results is the open question as to whether single chiral molecules can act as spin polarisers at all.
This intriguing incertitude is fortified by a recent report on the apparent absence of the CISS effect in single-molecule break junction experiments \cite{jacs_147_25043}.

The work presented here was stimulated by the delineated ambiguous picture of the CISS effect at the single-molecule level.  
The studies were conceived in a manner that avoids ambiguities and that reduces the complexity to a minimum.
To this end, a racemic mixture of heptahelicene (C$_{30}$H$_{18}$, [7]H) molecules was adsorbed as a monomolecular layer on Pb(111)\@.
The enantiomers $\Lambda$-[7]H and $\Delta$-[7]H self-assemble in chiral domains and are clearly discriminated in scanning tunnelling microscope (STM) images with intramolecular structure.
Spin selectivity of electron transport across the enantiomer is probed via a superconducting Pb tip functionalised with a magnetic Mn cluster.
The latter causes spin-polarised Yu-Shiba-Rusinov (YSR) states \cite{ptp_40_435,jetp_9_85,aps_21_75} within the Bardeen-Cooper Schrieffer energy (BCS) gap \cite{pr_108_1175} of the tip, which serve as sensitive probes for the spin polarisation of the tunnelling current.
Exploring spectra of the differential conductance ($\text{d}I/\text{d}V$, $I$: current, $V$: sample voltage) acquired atop the two chiral variants reveals a significant difference in the YSR signal strength and, thus, demonstrates the presence of the single-molecule CISS effect.
In addition, the identified dependence of the spectral signal on the current direction favours the enantiomers acting as spin polarisers rather than as spin filters. 

\section*{Enantiospecific quasielectron transport}

\begin{figure}
\centering
\includegraphics[width=0.95\textwidth]{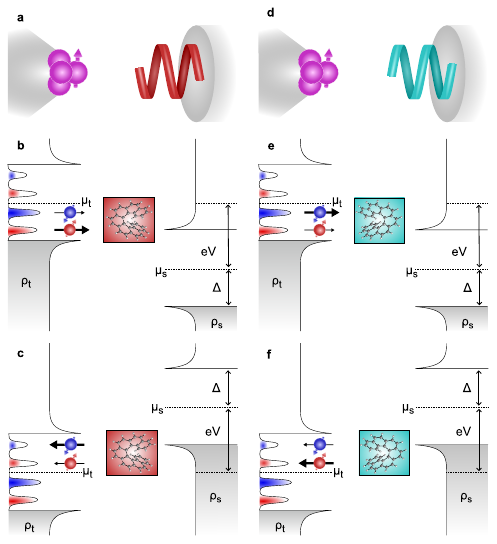}
\caption{\textbf{Spin-selective electron transport across enantiomers using superconducting electrodes.}
(a) Illustration of the tunnelling junction comprising a Mn-terminated Pb tip and a $\Lambda$-[7]H enantiomer on Pb(111)\@.
(b) Left: DOS of the tip ($\varrho_\text{t}$) showing the BCS energy gap of width $2\Delta$ and two pairs of spin-polarised YSR states.
Right: DOS of the sample ($\varrho_\text{s}$)\@.
Quasielectron transport with indicated spin occurs from tip to sample due to positive sample voltage $V=(\mu_\text{t}-\mu_\text{s})/\text{e}>0$\@.
(c) As (b) for $V<0$\@.
(d)--(f) As (a)--(c) for the $\Delta$-[7]H enantiomer.
The thickness of the horizontal arrows indicates the quasielectron transmission with the respective spin component.}
\label{fig1}
\end{figure}

Before presenting the experimental data it is appropriate to illustrate the charge transport across enantiomers in the presence of superconducting electrodes and YSR states (Fig.\,\ref{fig1}) because this junction composition has not been used to date for exploring the CISS effect.

Figures \ref{fig1}a--c depict the situation for the $\Lambda$-enantiomer.
The density of states (DOS) of the tip ($\varrho_\text{t}$) exhibits the BCS energy gap with symmetric flanks at $\mu_\text{t}\pm\Delta$ ($\mu_\text{t}$: chemical potential of the tip, $2\Delta$: width of the BCS energy gap) together with two pairs of electron-hole-symmetric intragap states.
The choice of two pairs is motivated by the actual experimental situation to be discussed below.
Owing to the functionalisation of the tip with a magnetic cluster, the YSR states are fully spin-polarised \cite{sciadv_7_eabd7302,nanoscale_14_15111,natcommun_15_459}.
Without loss of generality, the YSR state with high (low) binding energy $\varepsilon_1$ ($\varepsilon_0$) is assumed to exhibit spin-up (spin-down) polarisation, and the $\Lambda$-enantiomer to preferentially transmit spin-up quasielectrons at positive sample voltage $V=(\mu_\text{t}-\mu_\text{s})/\text{e}>0$ (e: elementary charge), i.\,e., for quasielectron tunnelling from the occupied YSR state to empty quasielectron states of the sample (Fig.\,\ref{fig1}b)\@.
At $T=0$, quasielectron transport sets in at $V=(\Delta+\varepsilon_{0,1})/\text{e}$.
Reversing the bias voltage polarity entails the inversion of the current direction and, therefore, the preferential transmission of spin-down electrons from the sample to the unoccupied YSR state with spin-down polarisation by the $\Lambda$-enantiomer (Fig.\,\ref{fig1}c)\@.
The opposite behaviour applies to the $\Delta$-enantiomer (Figs.\,\ref{fig1}d--f) with spin-down and spin-up polarisation of quasielectrons at, respectively, $V>0$ and $V<0$\@.
Therefore, it is expected that the two enantiomers interchange the signal strength among the two pairs of YSR states.

\section*{Adsorption and separation of a racemic mixture}

\begin{figure}
\centering
\includegraphics[width=0.8\textwidth]{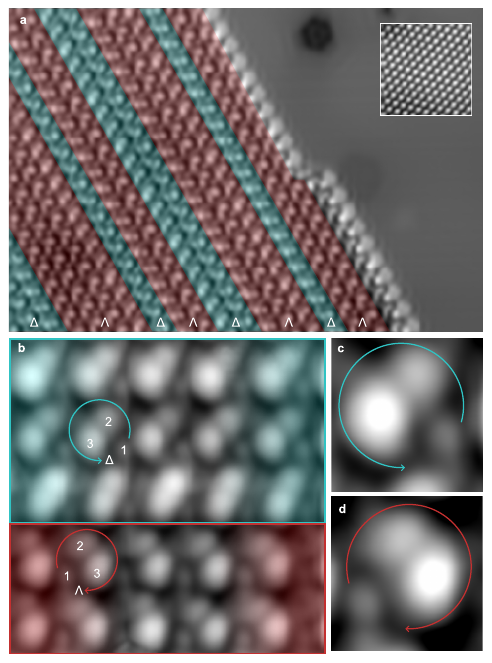}
\caption{\textbf{Separation of a racemic mixture of $\Lambda$-[7]H and $\Delta$-[7]H on Pb(111)\@.}
(a) Constant-current STM image of a [7]H island with marked enantiopure domains ($1\,\text{V}$, $50\,\text{pA}$, $30\,\text{nm}\times 20\,\text{nm}$)\@.
Inset: atomically resolved Pb(111) lattice ($10\,\text{mV}$, $50\,\text{pA}$, $4\,\text{nm}\times 4\,\text{nm}$)\@.
(b) Close-up view of a [7]H island with enantiopure domains indicated by shaded areas and individual $\Lambda$ and $\Delta$ molecules.
Numbers $1$--$3$ indicate a submolecular pattern with increasing apparent height (arc with arrow) reflecting the molecular helix with right (top) and left (bottom) handedness ($1\,\text{V}$, $50\,\text{pA}$, $6\,\text{nm}\times 6\,\text{nm}$)\@.
(c),(d) STM images of a single (c) $\Delta$-[7]H and (d) $\Lambda$-[7]H molecule embedded in the island ($1\,\text{V}$, $50\,\text{pA}$, $1.4\,\text{nm}\times 1.4\,\text{nm}$)\@.}
\label{fig2}
\end{figure}

Deposition of a racemic mixture of $\Lambda$-[7]H and $\Delta$-[7]H on Pb(111) at room temperature gives rise to single-layer islands with an ordered molecular superstructure (Fig.\,\ref{fig2}a) that is incommensurate with the substrate lattice (Supplementary Note S1)\@.
In contrast to observations from Cu(111) \cite{cgd_8_1890}, and the (111) surfaces of Ag and Au \cite{jpcc_118_29135,chir_32_975} where zigzag rows of alternating enantiomers form, on Pb(111) the molecules arrange into straight enantiopure rows and domains, which with $\langle 1\bar{1}0\rangle$ directions of the Pb(111) lattice (inset to Fig.\,\ref{fig2}a) subtend an angle of $(14\pm 2)^\circ$\@.
Enantiopure domains are a few molecules wide and are separated by molecular rows that appear with brighter STM contrast than within the domains, possibly due to different tilt angles of the molecular helical axes \cite{ape_18_015502}.
On-surface chiral recognition of enantiomers as encountered here is known from early atomic force microscope studies of polymers \cite{nature_368_440}.

It is essential to the spectroscopic findings of the present work to unambiguously identify the enantiomers in STM images.
The detailed view of a molecular island and individual enantiomers embedded therein (Figs.\,\ref{fig2}b--d) serve this purpose.
Figure \ref{fig2}b shows an STM image where a $\Lambda$-domain (bottom) adjoins a $\Delta$-domain (top)\@.
The spatial resolution enables the access to submolecular structure denoted as $1$, $2$, and $3$, which is assigned to the charge density of the lowest unoccupied frontier orbital (Supplementary Fig.\,S1)\@.
Such intramolecular resolution was best achieved for elevated voltages, which matches spectroscopy of $\text{d}I/\text{d}V$ atop the enantiomers where fingerprints of the frontier orbital appear for $V\geq 1\,\text{V}$ (Supplementary Fig.\,S2)\@.
From Fig.\,\ref{fig2}b as well as from the close-up views in Figs.\,\ref{fig2}c,d, the opposite sense of rotation towards increasing apparent height from $1$ to $3$ is evident for the two enantiomers.
The increasing apparent height from $1$ to $3$ is interpreted as the consequence of the acute angle between the molecular helical axis and the surface normal.

The resolution of intramolecular patterns does not only permit the unambiguous assignment of the enantiomer to the spectral quasielectron transport characteristics.
It additionally allows to explore possible variations of the enantioselective transmission at the atomic length scale.
Both aspects will be discussed in the following.

\section*{Spectroscopy with a spin-sensitive probe}

\begin{figure}
\centering
\includegraphics[width=0.9\textwidth]{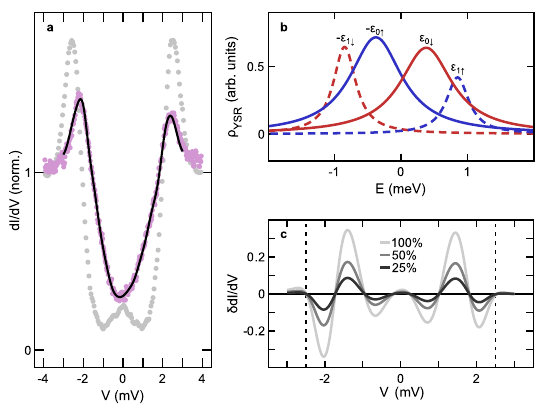}
\caption{\textbf{Expected effect of spin-selective quasielectron transport across enantiomers probed by a Mn-functionalised Pb tip.}
(a) Spectrum of $\text{d}I/\text{d}V$ recorded with a Mn-terminated tip atop clean Pb(111) in an external magnetic field of $B=40\,\text{mT}$\@.
The solid line is a fit to the Mn tip spectrum.
For comparison, $\text{d}I/\text{d}V$ data obtained with a clean Pb tip at $B=0$ are added as dots.
The spectra are normalised to unity at $5\,\text{mV}$\@.
Feedback loop parameters: $1\,\text{V}$, $50\,\text{pA}$\@.
(b) Lorentzian DOS ($\varrho_\text{YSR}$) used for describing the YSR states in the Mn tip spectrum in (a) ($\varepsilon_{i\sigma}$: YSR energy with respect to $E_\text{F}\equiv 0$, $i\in\{0,1\}$, spin $\sigma\in\{\uparrow,\downarrow\}$)\@.
(c) Simulated difference spectra $\delta\,\text{d}I/\text{d}V$ of the two enantiomers with indicated polarisation.
The dashed lines indicate the position of coherence peaks in the case of a clean Pb--Pb tunnelling junction.}
\label{fig3}
\end{figure}

To experimentally prove the electron spin polarisation in charge transport across the enantiomers, a spin-sensitive probe is required (Fig.\,\ref{fig1}), which is achieved by attaching a Mn cluster to the Pb tip.
The termination of the tip with a Mn cluster proceeded via the indentation of the tip apex into an island of Mn atoms coadsorbed with the molecules (see Methods for details) and was monitored by $\text{d}I/\text{d}V$ spectra of the clean Pb(111) surface (Fig.\,\ref{fig3}a)\@.
Deviating from the spectrum recorded with a clean Pb tip (grey dots in Fig.\,\ref{fig3}a) the spectrum acquired with the Mn-terminated tip exhibits an apparently reduced gap width, which is defined by a pair of peaks at $-2.26\,\text{mV}$ and $2.35\,\text{mV}$ (rather than at $\pm 2.52\,\text{mV}$ resulting from a clean Pb tip)\@.
These spectral changes are attributed to the presence of intragap states.
Indeed, the experimental $\text{d}I/\text{d}V$ data are best described (solid line in Fig.\,\ref{fig3}a) by assuming two pairs of YSR states represented each by a Lorentzian (see Methods for details) at energy $\varepsilon_0=\pm 0.38\,\text{meV}$ and $\varepsilon_1=\pm 0.85\,\text{meV}$ (Fig.\,\ref{fig3}b)\@.
An external magnetic field $B=40\,\text{mT}<B_\text{c}$ ($B_\text{c}=42\,\text{mT}$: critical field of Pb at $5\,\text{K}$ \cite{rmp_26_277}) was applied along the surface normal to ensure the spin polarisation of the YSR states \cite{science_358_772,prl_119_197002,sciadv_7_eabd7302}.
However, the effect to be discussed now is likewise present in the case of $B=0$, which demonstrates that the Mn cluster at the tip apex is large enough to stabilise its own magnetic moment.
Indeed, Mn clusters of as few as five atoms were shown to exhibit ferromagnetic order \cite{jcp_76_5636}.

\begin{figure}
\centering
\includegraphics[width=0.6\textwidth]{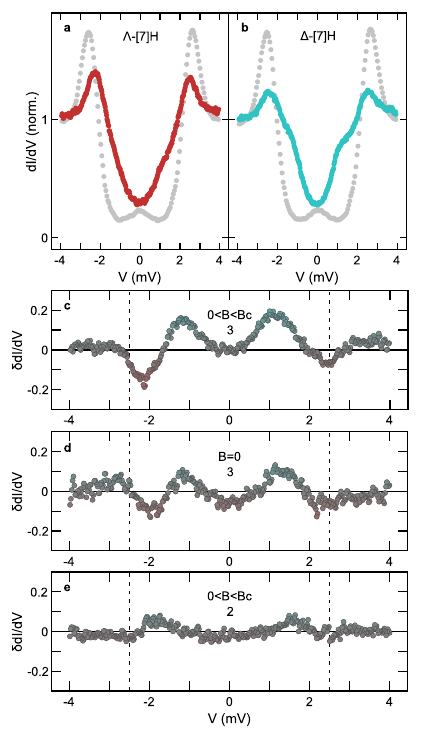}
\caption{\textbf{Experimental evidence for the single-enantiomer CISS effect.}
(a),(b) Spectra of $\text{d}I/\text{d}V$ obtained at $B=40\,\text{mT}$ with the same Mn-terminated tip as in Fig.\,\ref{fig3}a on the apparently highest part (site $3$) of (a) $\Lambda$-[7]H and (b) $\Delta$-[7]H\@.
Grey dots depict spectroscopic data recorded with a clean Pb tip.
The spectra are normalised to unity at $5\,\text{mV}$\@.
Feedback loop parameters: $1\,\text{V}$, $50\,\text{pA}$\@.
(c),(d) Difference spectra of (a) and (b) at (c) $B=40\,\text{mT}$ and (d) $B=0$\@.
(e) Difference spectrum obtained from $\text{d}I/\text{d}V$ data of $\Lambda$-[7]H and $\Delta$-[7]H acquired atop site $2$\@.}
\label{fig4}
\end{figure}

Prior to discussing the spectroscopic data obtained with the same tip on the enantiomers, the expected result is simulated (see Methods for details)\@.
Figure \ref{fig3}c depicts the difference of calculated $\text{d}I/\text{d}V$ spectra of the enantiomers, i.\,e., $\delta\,\text{d}I/\text{d}V=\text{d}I_\Delta/\text{d}V-\text{d}I_\Lambda/\text{d}V$, under the assumption of different degrees of polarisation, where $100\,\%$ ($0\,\%$) refers to the perfect transmission (suppression) of quasielectrons with a single spin component.
Qualitatively, the $\delta\,\text{d}I/\text{d}V$ curves are similar.
They exhibit pronounced minima close to the coherence peak positions (dashed lines) with $\delta\,\text{d}I/\text{d}V<0$ and maxima inside the BCS energy gap with $\delta\,\text{d}I/\text{d}V>0$\@.
This evolution reflects the preferential transmission of quasielectrons by the $\Delta$-enantiomer to the low-energy ($\varepsilon_0$) pair of spin-polarised YSR states compared to the $\Lambda$-enantiomer, whose transmission in turn is superior for the high-energy ($\varepsilon_1$) pair of YSR states with opposite spin polarisation.
Allowing in the simulations for reduced polarisations ($50\,\%$, $25\,\%$ in Fig.\,\ref{fig3}c) is seen to quench the $\delta\,\text{d}I/\text{d}V$ values.
In particular, the variations of $\delta\,\text{d}I/\text{d}V$ close to $V=0$ are attenuated.
Importantly, the pairs of minima and maxima of $\delta\,\text{d}I/\text{d}V$ occur each at symmetric positive and negative voltages, as expected from the CISS picture delineated in Fig.\,\ref{fig1}.

Figures \ref{fig4}a,b show the $\text{d}I/\text{d}V$ spectra of the two enantiomers obtained with the same Mn-terminated Pb tip as used for the spectrum in Fig.\,\ref{fig3}a atop the brightest part of the molecules in STM images ($3$ in Fig.\,\ref{fig2}b)\@.
At the feedback loop parameters used for spectroscopy, the apparent heights of $\Lambda$-[7]H and $\Delta$-[7]H differ by less than $5\,\text{pm}$, which entails nearly the same tip--enantiomer distances.
The plots also include as grey dots molecule spectra acquired with a clean Pb tip.
They are essentially indistinguishable from spectra of pristine Pb(111), which, importantly, shows that the adsorbed enantiomers themselves do not induce intragap states, i.\,e., do not act as magnetic impurities as previously observed for $\alpha$-helix polyalanine on NbSe$_2$ \cite{nl_19_5167}.
In addition, a possible change in the order parameter of the conventional superconductor due to the presence of chiral molecules \cite{njp_18_113048} can be excluded.
Using now the experimental data for calculating $\delta\,\text{d}I/\text{d}V$ yields the evolution presented in Fig.\,\ref{fig4}c for $B=40\,\text{mT}$, which is remarkably similar to the expectations (Fig.\,\ref{fig3}c)\@.
In the field-free case (Fig.\,\ref{fig4}d), the excursions of $\delta\,\text{d}I/\text{d}V$ to positive and negative values are attenuated, but the general trend remains invariant.
These spectroscopic results are the main findings of the studies presented here.
They demonstrate the CISS effect at the single-enantiomer level.

The good agreement of experimental and simulated $\delta\,\text{d}I/\text{d}V$ traces corroborates the dependence of the CISS effect on the current direction (Fig.\,\ref{fig1}) and, therefore, qualifies the enantiomers as spin polarisers rather than as spin filters \cite{acsnano_19_42046}.
Additionally, it is conjectured here, that the spin polarisation is generated within the respective enantiomer alone.
Indeed, a spin polarisation of the substrate electron DOS caused by the interaction with the adsorbed chiral molecules \cite{nl_19_5167} can be excluded because the symmetry in $\delta\,\text{d}I/\text{d}V$ traces (Figs.\,\ref{fig4}c,d) contrasts the quasielectron transport behaviour expected from a spin-polarised substrate electron DOS (Supplementary Fig.\,S3)\@.
Furthermore, the spatially resolved spectroscopy experiments unveiled that the extent of the CISS effect depends on the actual site of the enantiomer, which is evidenced by $\delta\,\text{d}I/\text{d}V\approx 0$ obtained at position $2$ in the relevant voltage range (Fig.\,\ref{fig4}e)\@.
Because $2$ is apparently lower than $3$, where the maximum effect is observed (Figs.\,\ref{fig4}c,d), the result in Fig.\,\ref{fig4}e reflects the previously reported dependence of the spin polarisation on the length of the chiral molecule \cite{pccp_15_18357,jpcc_119_14542,jpcc_123_3024,jpcc_124_11716}.
The spatial variability of the CISS effect likewise offers an explanation to the null result in recent break junction experiments \cite{jacs_147_25043}, where the lacking spatial resolution requires the averaging over a multitude of current-voltage characteristics obtained from contacts without direct access to the molecule orientation.

Additional experiments were performed to exclude, e.\,g., a possible dependence of the effect on the enantiomer adsorption site.
Because of the incommensurate adsorption structure, adjacent molecules of the same enantiopure domain occupy different sites of the Pb(111) surface and may therefore exhibit different spectra.
However, Supplementary Fig.\,S4 demonstrates the absence of spectral changes due to the adsorption site; rather, the $\text{d}I/\text{d}V$ spectra reflect the spin selectivity caused by the chirality alone.
Another important aspect is the exclusion of electrostatic effects in the observations.
To this end, the spectra of the molecules were explored for different tip--enantiomer distances (Supplementary Fig.\,S5)\@.
The presence of the tip locally modifies the surface potential with implications on band bending and work function changes \cite{acsnano_18_6028} as well as on molecular electric dipoles \cite{natcommun_7_10744,chemsci_15_14905,jacs_147_36453}.
Because the spectra remain invariant for a large range of tip--molecule separations, such effects are considered irrelevant here.
Finally, a possible influence of spin-orbit coupling (SOC) due to the presence of Pb is noteworthy because the enantiomers could then take the role of orbital filters, which would entail spin filtering due to SOC in the metal electrodes \cite{natmater_20_638}.  
Calculations of the Pb bulk band structure with and without SOC \cite{prb_84_115144} unveiled, however, that the strongest SOC-induced changes occur for bands with energies far above the Fermi level at wave vector $k=0$\@.  
In addition, SOC affects bands crossing $E_\text{F}$ only for elevated $k$.  
These findings for bulk Pb are in agreement with more recent experimental and simulated band structure data for Pb(111) \cite{scirep_13_1689}.  
The results therefore support the irrelevance of SOC in the present spectroscopy experiments, where quasielectron transport is studied in the vicinity of $E_\text{F}$ and where tunnelling with $k\neq 0$ is exponentially suppressed.

While the reliable interpretation of the experimental data relies on the difference spectra and, thus, on the invariance of the tip structural integrity in spectroscopy atop various enantiomers, different Mn-terminated tips with different YSR resonance textures were independently prepared in order to verify the robustness of the CISS effect.
The different tips gave rise to altered YSR states that yielded modified $\delta\,\text{d}I/\text{d}V$ data.
However, the basic trend of $\delta\,\text{d}I/\text{d}V$ (Fig.\,\ref{fig3}c, Figs.\,\ref{fig4}c,d), which is indicative of the CISS effect, was retained (Supplementary Figs.\,S6--S8)\@.

\section*{Conclusions and outlook}

The use of spin-polarised YSR states as probes for the quasielectron spin state enables the unambiguous evidence for the CISS effect in single-enantiomer STM junctions with superconducting electrodes.
The reported experimental approach to the CISS effect reported here avoids the anchoring of the chiral molecules to a ferromagnetic substrate as well as the magnetisation reversal for detecting magnetoresistance in the charge transport studies.
Consequently, electrostatic effects, such as work function changes or electric-dipole modifications, are excluded.
The dependence of the spin-selective quasielectron transport on the current direction demonstrates the spin polariser behaviour of the individual enantiomers.
The spatially resolved spectroscopy indicates the impact of the effective molecular length on the spin polarisation and unveils the intramolecular variability of the CISS effect, which in spatially averaging transport experiments may feign the absence of the single-enantiomer spin polariser.

Based on the presented findings we foresee reinforced computational efforts in understanding the single-enantiomer CISS effect.
Moreover, the simple model system character allows for artificial atomic-scale modifications of the enantiomer environment, such as the presence of magnetic impurities or quantum-confined electron states, which will further help understand the CISS effect at the ultimate size.

\section*{Experimental methods}

The experiments were carried out with an STM operated in ultrahigh vacuum (base pressure $5\cdot 10^{-9}\,\text{Pa}$) and at low temperature ($5\,\text{K}$)\@.
Clean crystalline (111) surfaces of Pb were prepared by Ar$^+$ ion bombardment and annealing.
Chemically etched W wire (purity $99.95\,\%$) was used as the tip material and further treated by field emission on the surface of a Au crystal. 
Superconducting tips were obtained by coating the tip apex with Pb substrate material in the course of performing field emission with the Au-coated W tip on a pristine Pb substrate. 
Manganese was deposited on clean Pb(111) at room temperature with low coverage using electron beam heating of purified ($99.5 \%$) Mn powder inside a crucible. 
The termination of Pb-coated tips with Mn proceed by indentations of the tip apex into adsorbed Mn clusters.
The functionalisation was further optimized by single-atom transfer from the tip to the surface \cite{prl_94_126102,jpcm_20_223001}.  
Heptahelicene molecules were sublimated from the solid phase in a heated ($370\,\text{K}$) Ta crucible and deposited on Mn-precovered Pb(111) at room temperature.
Raw constant-current and constant-height STM image data were further processed with the Nanotec Electronics WSxM software \cite{rsi_78_013705}.
Spectroscopy of $\text{d}I/\text{d}V$ proceeded via the sinusoidal modulation ($100\,\mu\text{V}_{\text{rms}}$, $360\,\text{Hz}$) of the dc bias voltage and measuring the first harmonic of the ac current response of the junction with a lock-in amplifier.
The voltage was applied to the sample giving rise to unoccupied (occupied) sample states at $V>0$ ($V\leq 0$)\@.

\section*{Computational methods}

To characterize the Mn-terminated Pb tip, the number of induced YSR resonance pairs is desirable.
Spectra of $\text{d}I/\text{d}V$ acquired with the Mn tip above clean Pb(111) were therefore fit as follows.
The tunnelling current is modeled as
\begin{equation}
I\propto\int\varrho_\text{t}(E)\varrho_\text{s}(E-\text{e}V)[f_\text{t}(E)-f_\text{s}(E-\text{e}V)]\,\text{d}E
\label{eq:i}
\end{equation}
with $E$ the energy of the tunnelling electron, $f_\text{t,s}=1/\{\exp[\beta(E-\mu_\text{t,s})]+1\}$ the Fermi-Dirac distribution function of tip and sample ($\beta=1/(\text{k}_\text{B}T)$ with $\text{k}_\text{B}$ the Boltzmann constant),
\begin{equation}
\varrho_\text{s}=\left\{
\begin{array}{ll}
\text{sign}(E)\cdot\Re\left[\dfrac{E-\text{i}\gamma}{\sqrt{(E-\text{i}\gamma)^2-\Delta^2}}\right] & \vert E\vert>\Delta \\
0 & \vert E\vert\leq\Delta
\end{array}
\right.
\label{eq:doss}
\end{equation}
the Bogoliubov quasielectron DOS of the sample ($\text{sign}(E)=\pm 1$ for $E>\mu_\text{s}$ ($+$) and $E<\mu_\text{s}$ ($-$), $\Re$: real part, $\gamma$: Dynes parameter reflecting the finite quasielectron lifetime \cite{prl_41_1509}, $\text{i}^2=-1$), and
\begin{equation}
\varrho_\text{t}(E)=a_0L_0(E)+a_1L_1(E)+\left\{
\begin{array}{ll}
\text{sign}(E)\cdot\Re\left[\dfrac{E-\text{i}\gamma}{\sqrt{(E-\text{i}\gamma)^2-\Delta^2}}\right] & \vert E\vert>\Delta \\
0 & \vert E\vert\leq\Delta
\end{array}
\right.
\label{eq:dost}
\end{equation}
the Bogoliubov quasielectron DOS of the tip, which contains two Lorentzians of the form
\begin{equation}
L_i(E)=\frac{1}{\pi}\frac{\Gamma_i}{(E-\varepsilon_i)^2+\Gamma_i^2}\quad (i\in\{0,1\})
\label{eq:lor}
\end{equation}
($\varepsilon_i$ and $\Gamma_i$ denote, respectively, energy and full width at half maximum of the Lorentzian)\@.
The tunnelling transmission factor is omitted from equation (\ref{eq:i}) because the energies of tunnelling quasielectrons are much lower than the work functions of the electrodes, which justifies the assumption of an essentially constant tunnelling transmission.
The numerical derivative of equation (\ref{eq:i}) is convoluted with
\begin{equation}
\chi_\text{m}=\left\{
\begin{array}{ll}
\dfrac{2}{\pi}\cdot\dfrac{\sqrt{V_\text{m}^2-V^2}}{V_\text{m}^2} & \vert V\vert\leq V_\text{m} \\
0 & \vert V\vert>V_\text{m}
\end{array}
\right.
\label{eq:mod}
\end{equation}
in order to consider spectroscopic broadening due to the finite sinusoidal modulation amplitude $V_\text{m}$ and then fit to the experimental $\text{d}I/\text{d}V$ data.

The net tunnelling current between tip and sample is written as:
\begin{equation}
I=I_{\text{t}\rightarrow\text{s}}-I_{\text{s}\rightarrow\text{t}}
\end{equation}
where $I_{\text{t}\rightarrow\text{s}}$ ($I_{\text{s}\rightarrow\text{t}}$)  describes the tunnelling from occupied states of the tip (sample) into unoccupied states of the sample (tip), i.\,e.,
\begin{eqnarray}
I_{\text{t}\rightarrow\text{s}}(\text{e}V)&=&\int\varrho_\text{t}(E-\text{e}V)f_\text{t}(E-\text{e}V)\varrho_\text{s}(E)[1-f_\text{s}(E)]\,\text{d}E \\
I_{\text{s}\rightarrow\text{t}}(\text{e}V)&=&\int\varrho_\text{t}(E-\text{e}V)[1-f_\text{t}(E-\text{e}V)]\varrho_\text{s}(E)f_\text{s}(E)\,\text{d}E
\end{eqnarray}
Identifying the tip DOS as a sum of two spin channels ($\varrho_\text{t}=\varrho^{\uparrow}_\text{t}+\varrho^{\downarrow}_\text{t}$) the total current is now composed of the four spin-resolved contributions
\begin{equation}
I=I^{\uparrow}_{\text{t}\rightarrow\text{s}}+I^{\downarrow}_{\text{t}\rightarrow\text{s}}-I^{\uparrow}_{\text{s}\rightarrow\text{t}}-I^{\downarrow}_{\text{s}\rightarrow\text{t}}
\end{equation}
To model the spin-polarising capability of the molecules, the allowed spin channels are associated with the handedness of the individual enantiomers. 
Retaining the convention of Fig.\,\ref{fig1}, enantiomer $\Lambda$-[7]H only transmits spin-up (spin-down) quasielectrons from tip to sample (sample to the tip)\@. 
Therefore, assuming perfect transmission implies $I^{\downarrow}_{\text{t}\rightarrow\text{s}}=0=I^{\uparrow}_{\text{s}\rightarrow\text{t}}$ for $\Lambda$-7[H] and $I^{\uparrow}_{\text{t}\rightarrow\text{s}}=0=I^{\downarrow}_{\text{s}\rightarrow\text{t}}$ for $\Delta$-[7]H\@.

Imperfect polarisation can likewise be modeled by introducing a phenomenological parameter $\alpha$ with $0\leq\alpha\leq 1$\@.
A perfect polariser is described by $\alpha=1$, whereas $\alpha=0$ implies the absence of the spin-polarising effect.
The currents across the enantiomers then read
\begin{eqnarray}
I_{\Lambda}&=& \frac{1}{2} \left[ (1+\alpha) I^{\uparrow}_{\text{t}\rightarrow\text{s}}+(1-\alpha)I^{\downarrow}_{\text{t}\rightarrow\text{s}}-(1-\alpha)I^{\uparrow}_{\text{s}\rightarrow\text{t}}-(1+\alpha) I^{\downarrow}_{\text{s}\rightarrow\text{t}} \right] \\
I_{\Delta}&=& \frac{1}{2} \left[ (1-\alpha)I^{\uparrow}_{\text{t}\rightarrow\text{s}}+(1+\alpha) I^{\downarrow}_{\text{t}\rightarrow\text{s}}-(1+\alpha) I^{\uparrow}_{\text{s}\rightarrow\text{t}}-(1-\alpha)I^{\downarrow}_{\text{s}\rightarrow\text{t}}\right]
\end{eqnarray}
The simulated $\delta\,\text{d}I/\text{d}V=\text{d}I_{\Delta}/\text{d}V-\text{d}I_{\Lambda}/\text{d}V$ signal reads
\begin{equation}
\delta\,\frac{\text{d}I}{\text{d}V}=\alpha\left(\frac{\text{d}I^{\downarrow}_{\text{t}\rightarrow\text{s}}}{\text{d}V}+\frac{\text{d}I^{\downarrow}_{\text{s}\rightarrow\text{t}}}{\text{d}V}-\frac{\text{d}I^{\uparrow}_{\text{t}\rightarrow\text{s}}}{\text{d}V}-\frac{\text{d}I^{\uparrow}_{\text{s}\rightarrow\text{t}}}{\text{d}V}\right)
\label{eq:delta}
\end{equation}
which exhibits maximal variations for $\alpha=1$ and vanishes for $\alpha=0$\@.

A spin-polarised current can likewise be achieved by the imbalance of majority and minority electrons in the case of spin-polarised DOS of tip and sample according to the Jullière model \cite{pla_54_225}.
However, $\delta\,\text{d}I/\text{d}V$ for spin-polarised DOS markedly differs from $\delta\,\text{d}I/\text{d}V$ resulting from the dependence of the spin polarisation on the current direction (Supplementary Figs.\,S3a,b)\@.
Consequently, the agreement of experimental data (Fig.\,\ref{fig4}) with equation (\ref{eq:delta}) corroborates the interpretation in terms of the single-enantiomer CISS effect.

\section*{Acknowledgments}

Funding by the Deutsche Forschungsgemeinschaft (grant no.\ KR 2912/21-1), the Agence Nationale de la Recherche (grant no.\ ANR-23-CE09-0036-01), and the German Federal Ministry of Education and Research within the ''Forschungslabore Mikroelektronik Deutschland (ForLab)'' initiative is acknowledged. 
Discussions with Krisztián Palotás, Maximilian Kögler, and Karl Rothe are appreciated.

\section*{Supplementary information}  

Supplementary information is available at DOI: [hyperlink DOI] 

\section*{Competing interests}  
The authors declare no competing financial interests.

\section*{Author contributions}

L.\ M.\ and N.\ N.\ performed the experiments. 
L.\ M.\ and J.\ K.\ wrote the manuscript. 
N.\ N.\ and J.\ K.\ conceived the experiments.
All authors analyzed the data, discussed their interpretation, and commented on the draft.

\section*{Data availability}  

The presented data will be provided upon reasonable request to the authors.

%\bibliography{ref}
%

\end{document}